\documentclass{article}

\usepackage{graphicx}
\begin{document}

\title{The Phosphine Controversy:\\ Is it Phosphine? Is there life on Venus?}
\author{Priya Hasan}

\maketitle

\begin{abstract}
On 14th September 2020, the Royal Astronomical Society made an official statement coupled with a webminar on the discovery of phosphine on Venus. Single-line millimetre-waveband spectral detections of phosphine (with a signal-to-noise ratio of  $\approx$ 15$\sigma$) from the JCMT and ALMA telescopes indicated a phosphine abundance of 20 ppb\footnote{ppb: parts per billion}, 1000 times more than that on the Earth.
 Phosphine is an important biomarker and immediate speculation  in the media about indicators of life being found on Venus followed. This article presents an analysis of the study and the results on the observation of the spectral absorption feature of phosphine  in the clouds of Venus, thus implying as a potential biosignature. If phosphine is produced through biotic, as opposed to abiotic pathways, the discovery could imply a significant biomass  in  the  Venusian  atmosphere.  The discovery led to a major controversy with criticism of the analysis and results and responses to it. The issue remains unresolved, leading to a fresh interest in the study of Venus including ground-based observations as well as space-probes that can answer these questions conclusively. 

\end{abstract}


\begin{center}

\it{`We are alive and we resonate with idea of life elsewhere, but only careful accumulation and assessment of the evidence can tell us whether a given world is inhabited.' 
  \\
  Carl Sagan}
 \end{center}

\section*{Introduction}

Phosphine gas is not very pleasant. Pure phosphine is odourless, but technical grade samples smell like rotting fish, toxic and spontaneously flammable, with the chemical formula of $\rm{PH_3}$. 

On 14th September 2020, the Royal Astronomical Society made an official statement coupled with a webminar on the discovery of phosphine on Venus. At the heart of this announcement was the paper by \cite{greaves2020} titled `Phosphine gas in the cloud decks of Venus'. It was about the detection of phosphine in the temperate, but hyperacidic clouds of Venus with an abundance of 20 ppb. They also made a very detailed study of various chemical pathways of generation of phosphine in the given abundance by lightning, volcanoes or meteorites. They concluded, that the observed abundance of phosphine could  possibly be a result of biological processes. They also stated that a detection of statistical significance needed to be backed up by observations to validate the spectral features of phosphine and of the conditions in the Venusian atmosphere. 

Significant work in this context was done by \cite{bains2020} who at made a detailed study of all the possible chemical processes that can generate these amounts of phosphine. \cite{seager2020} proposed a Venusian aerial biosphere to explain the presence of phosphine as a possible hypothesis. 

\section{Biosignatures: Why Phosphine?}

Wikipedia defines biosignatures as:{ \it {A biosignature is any substance -- such as an element, isotope, or molecule -- or phenomenon that provides scientific evidence of past or present life. Measurable attributes of life include its complex physical or chemical structures and its use of free energy and the production of biomass and wastes}}. The obvious way forward is to identify molecules present on the Earth that could work as biosignatures. An ideal bio-signature should have its sole source from living organisms be  intrinsically strong and easily identifiable spectroscopically.

\begin{figure}[ht]
\caption{Oxygen as a biosignature: False positive mechanisms for oxygen on a variety of planet scenarios. The main contributors to a spectrum of the planet's atmosphere are shown in large rectangles. The yellow and red circles represent molecules that would confirm a false positive biosignature and and molecules crossed confirm  false positive biosignatures if not detected. Cartoon adapted from \cite{mead} (Image Credit: Wikipedia).}
\label{o2} 
\centering
\includegraphics[width=1.0\textwidth]{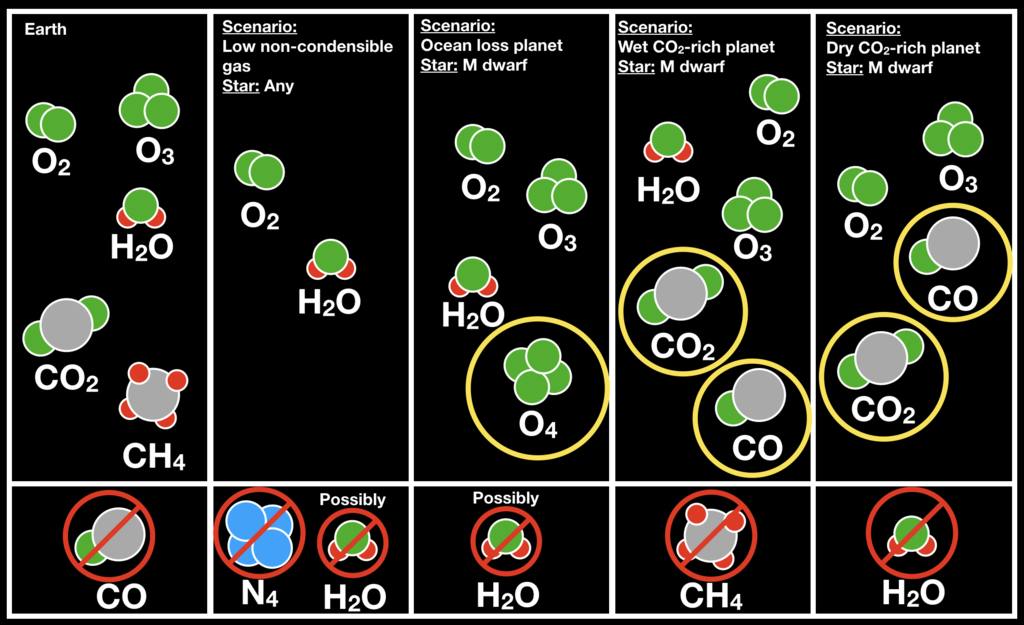}
\end{figure}

The element that comes first to the mind is oxygen. However, as shown by \cite{mead}, oxygen is not most suitable, as it creates several false positive mechanisms on a variety of planet scenarios are possible (see Fig.~\ref{o2}). This implies that detection as well as non-detection of certain molecules could lead to false positive biosignatures\footnote{A false positive is a set of non-biological processes that can mimic the detectable feature of a biosignature. False negative biosignatures occur  where life may be present on another planet, but potential biosignatures are  undetectable.}.

In the Earth's atmosphere, methane reacts with oxygen to get converted to carbon-dioxide and water, very efficiently. However, the atmosphere is found to have an abundance of 1 molecule per million and this can only be explained by a constant source of methane production (life) that oxygen cannot keep pace with in conversion to carbon-dioxide and water. Hence due to an imbalance of the production and destruction  of methane, we still find some of it in our atmosphere. This is an indicator of life in an oxygen rich environment. 

\begin{figure}[h]
\caption{Methane as a viable biosignature on the Earth. (Image Credit:  Wikipedia).}
\label{methe} 
\centering
\includegraphics[width=0.7\textwidth]{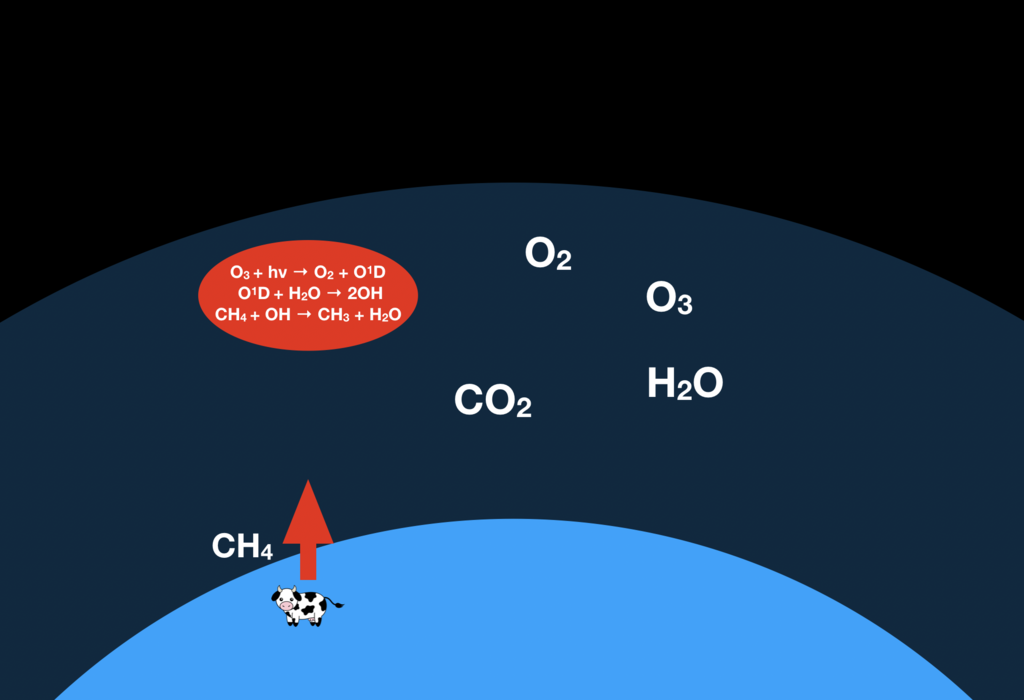}
\end{figure}

Biogenic methane production is the main contributor to the methane flux coming from the surface of Earth. Methane has a photochemical sink in the atmosphere, and hence can be detected only in the presence of biogenic methane production that is at a rate larger than its oxidation (Fig.~\ref{methe}).  \cite{arney} stated that the detection of  methane in the atmosphere of another planet can  a viable biosignature, especially if the host star is of $G$ or $K$ type.

\begin{figure}[h]
\caption{Methane as a biosignature on Mars: Generation and Destruction (Image Credit:  Wikipedia).}
\label{meth} 
\centering
\includegraphics[width=0.7\textwidth]{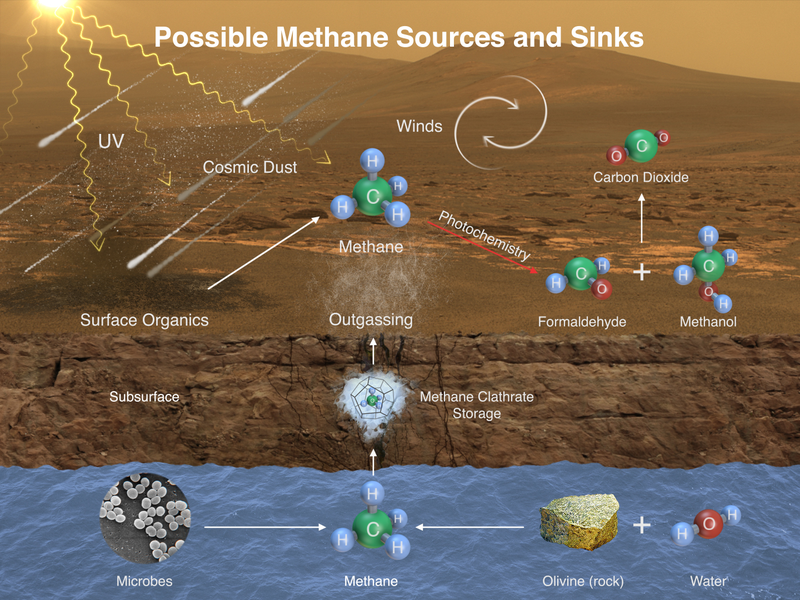}
\end{figure}

A similar argument was used to study methane as a biomarker on Mars. However, the processes  of generation and destruction of methane on Mars are not fully understood (See Fig.~\ref{meth}).  This is still controversial and a mass spectrometer measuring the isotopic proportions of carbon-12 to carbon-14 in methane is used to validate the bio-origin of methane \cite{zah}.


Phosphine is similar to ammonia, made of a single atom of phosphorous and three atoms of hydrogen. On the Earth, we have a phosphine abundance as 1 part per trillion. In an oxidizing environment, phosphine will immediately react with oxygen to form phosphorous acid or phosphoric acid. 
  
Phosphine is exclusively associated to life and is not generated by other natural processes involving the geology or atmosphere of our planet. Hence its presence in the oxidizing atmosphere, indicates a constant generation, caused by biological processes exceeding the rate of oxidation, making it a reliable biomarker.
 
On the Earth, phosphine is associated with anaerobic ecosystems, and as such it is a potential biosignature gas in anoxic exoplanets. It has also been found on the gas giants Jupiter and Saturn. These planets have a hydrogen rich atmosphere, and the phosphine gets generated in the high temperature and pressure environment getting mixed well with the gas layers due to convection. In the case of rocky planets, the surface provides a natural barrier stopping the phosphine produced at the core from getting mixed with the atmosphere. Hydrogen is more likely to combine with oxygen to form water, or carbon to form methane. Hence for phosphorous to combine with Hydrogen in its very low abundance, is very improbable. Hence the detection of phosphine on a temperate rocky planet, is a robust indicator of life \footnote{A high temperature could destroy the phosphine. Hence a process of regeneration of phosphine would be necessary for detection.}. On the Earth, phosphine could be produced directly by microbial reduction of more oxidized phosphorus species or indirectly by microbial production of reduced phosphorous compounds, such as hypophosphite leading to $\rm{PH_3}$, implying a bio-origin. 
  
An excess of phosphine would be due to an imbalance in the production and oxidation processes and can only be produced by life forms on rocky planets, for example, the Earth. If phosphine is produced through biotic, as opposed to abiotic pathways, the discovery could imply a significant biomass in the Venusian  atmosphere \cite{lingloeb}.
 
There are a few other mechanisms that can produce phosphine. Even if it was made in the lower layers, it would be destroyed before it reaches layers where it would be found. Volcanoes, thunder, etc. can produce it, but an appropriate hydrogen pressure would be required to keep the phosphine. A document of more than 100 pages was made to study all such mechanisms \cite{bains2020}. There is also strong ultraviolet (UV) absorbtion in the Venusian atmosphere. Venus was probably habitable in the past till the green house effect took over to make the planet inhabitable.  

Also, we should always keep in mind that life elsewhere could  be very different from life on Earth from a bio-chemical perspective, specially with the variety of bio-diversity on the Earth. 
 
\section{Venus: The Evil Twin}

Venus has been called the evil twin of the Earth since its size and mass is comparable. `Heaven and Hell' were the words used by Carl Sagan to describe the twins: Earth and Venus. The atmosphere of Venus is composed of 96.5\% $\rm{CO_2}$ and 3.5\% $\rm{N_2}$ and its present surface conditions are very different from that on Earth.

Life of Venus has been discussed in the past.  \cite{msagan,grin,cock} proposed that the conditions between the lower and middle atmosphere were conducive to (terrestrial) biology, while higher altitudes could possibly host microorganisms. Life in the Venusian cloud layers was also considered by \cite{schulz} due to the presence of conducive chemical and physical conditions, like the presence of sulfur compounds, carbon dioxide ($\rm{CO_2}$), and water, and moderate temperatures ($0 - 60^0$ C) and pressures ($\approx 0.4 - 2$ atm).

\begin{figure}[h]
\caption{The potential for microorganisms to survive in Venus' lower clouds and contribute to the observed bulk spectra in a hypothesis paper \cite{limaye} (Image credit: Limaye et al, doi: 10.1089/ast.2017.1783.)}
\label{lim} 
\centering
\includegraphics[width=10cm, height=7.0cm]{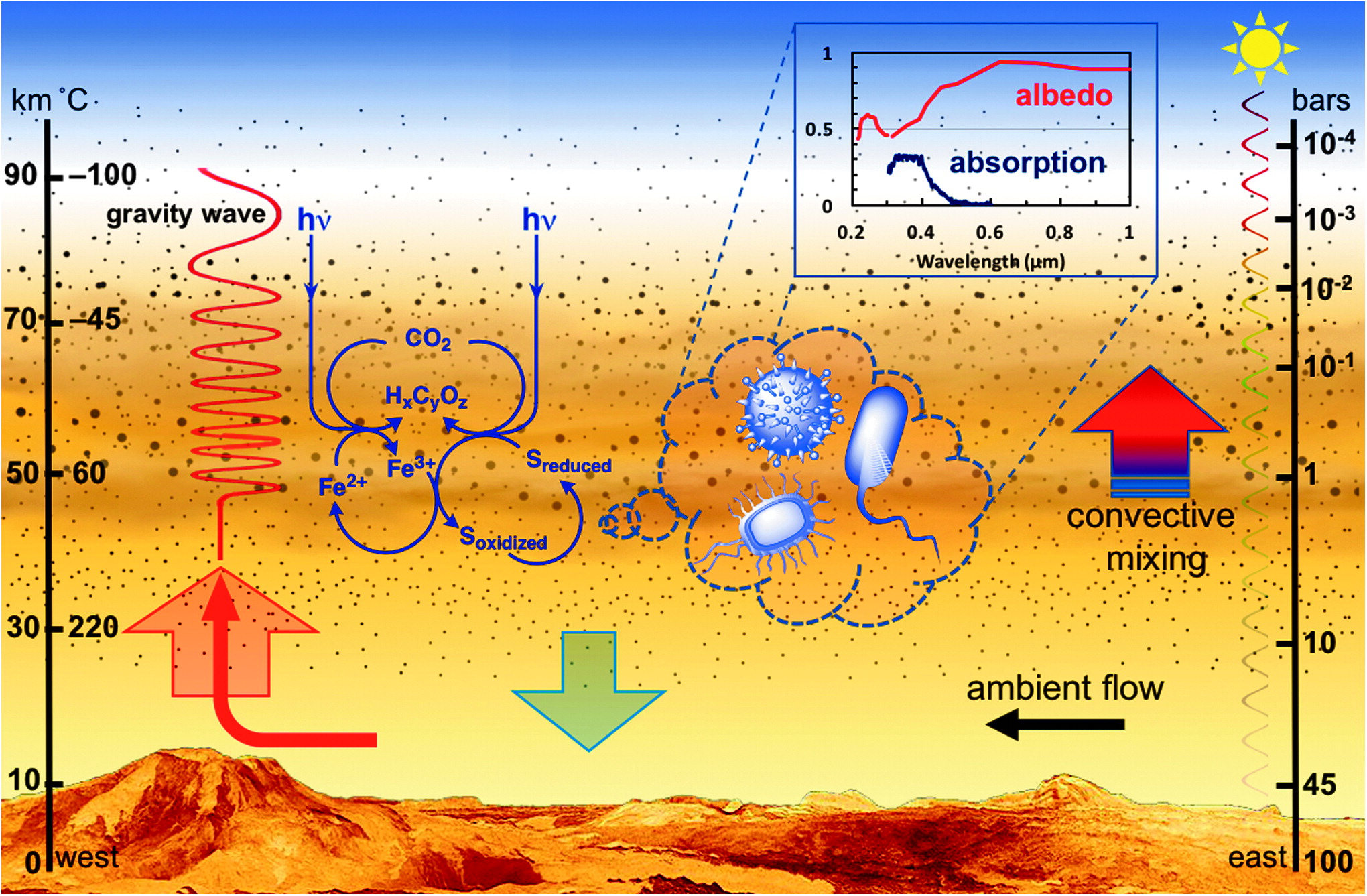}
\end{figure}

In an hypothesis article \cite{limaye}, proposed that the lower cloud layer of Venus ($47.5 - 50.5$ km) is an region favorable to microbial life, as it has moderate temperatures and pressures ($\approx $ $60^0$ C and 1 atm), and the presence of micron-sized sulfuric acid aerosols (Fig.~\ref{lim}).

The formidable Venusian surface has a blistering temperature of 470$^{0}$ C and a pressure 92 times higher than that at the surface of Earth. This is similar to the pressure found in the depth of 900 m of Earth's oceans. This is an extreme hostile environment for life, as we know it. The Russian Venera program was  a series of space probes between 1961 and 1984 to study Venus. Ten probes successfully landed on the surface of the planet, including the two Vega program and Venera-Halley probes, while thirteen probes successfully entered the Venusian atmosphere. However, due to the hostile surface conditions, the probes could only survive for  short periods ranging from 23 minutes to two hours. 

The surface of Venus is very hot ($\approx 400^0$ C) , but at a height of 45--60 km the temperature is about $30^0$ C and the pressure is 1 atm. This is possible region where the physical conditions are suitable for life, as we know it. It is believed that there could have been similar conditions (cooler and wetter) on the Earth and Venus at an earlier time and as conditions on Venus turned hostile, life could have migrated to this more temperate zone in the Venusian atmosphere. However, the chemical composition of the atmosphere is very different from that on the Earth, with clouds or droplets of sulphuric acid. Hence, life if it exists in these conditions, would be very different from we know it to be. 

\section{Data: JCMT and ALMA}
Phosphine as a probable biomarker hence seemed to be good candidate for detection on Venus.
Phosphine is made up of a single atom of phosphorous and three atoms of hydrogen. The atoms in the molecule can make transitions between different rotation energy levels corresponding to certain characteristic frequencies or wavelengths by absorbing solar radiation. One such transition from the $J=1 - 0$ levels is at a frequency of 266944.51 MHz = 1.123 mm. Observations of this spectral line can confirm the presence as well as the abundance of phosphine and the temperature and pressure of the region. 

\begin{figure}[!t]
\caption{Model of James Clerk Maxwell Telescope (JCMT). Photographed at the Royal Astronomical Society's National Annual Meeting 2009. (Image Credit: Wikipedia).}
\label{jcmt} 
\centering
\includegraphics[width=7cm, height=5cm]{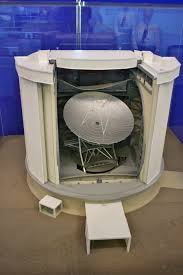}
\end{figure}

The James Clerk Maxwell Telescope (JCMT) is a submillimetre-wavelength radio telescope at Mauna Kea Observatory in Hawaii, USA. The telescope is near the summit of Mauna Kea at 4,092 m. Its primary mirror has a size of 15 m. It is the largest single-dish telescope that operates in submillimetre wavelengths of the electromagnetic spectrum (far-infrared to microwave) (Fig.~\ref{jcmt}).

\begin{figure}[!t]
\caption{Spectrum of PH$_3$(1-0) with JCMT \cite{greaves2020}}
\label{jcmts} 
\centering
\includegraphics[width=9cm, height=5cm]{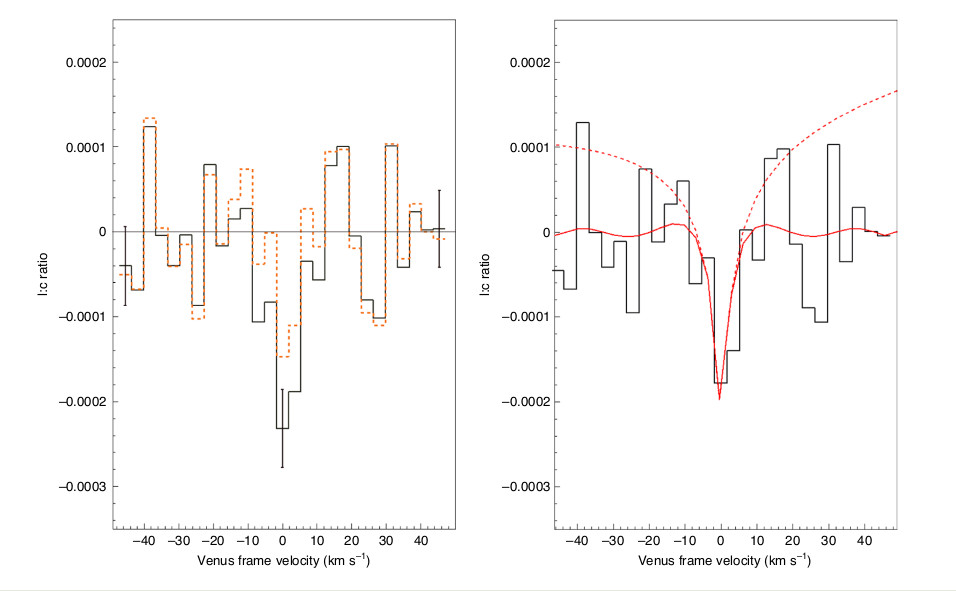}
\end{figure}

The Fig.~\ref{jcmts} shows the spectrum obtained from the JCMT. The spectral ripples\footnote{Spectral ripples are instrumental artefacts in the data that become apparent when observing an object as bright as Venus.} had to be fitted and removed to obtain the spectrum shown in the left panel with the residual line present inside velocity ranges of $v =  8$ km/s (solid, black) and $v = 2$ km/s (dashed, orange). The data was binned for clarity into histograms on the x axis; representative $1\sigma$ error bars are also shown. The right panel shows the adopted mid-range solution with $v = 5$ km/s (histogram), overlaid with our model for 20 ppb abundance by volume. The solid red curve shows this model after processing with the same spectral fitting as used for the data. 

\begin{figure}[!t]
\caption{Atacama Large Millimeter/submillimeter Array (ALMA)(Image Credit: P. Horálek/ESO)}
\label{alma} 
\centering
\includegraphics[width=7cm, height=5cm]{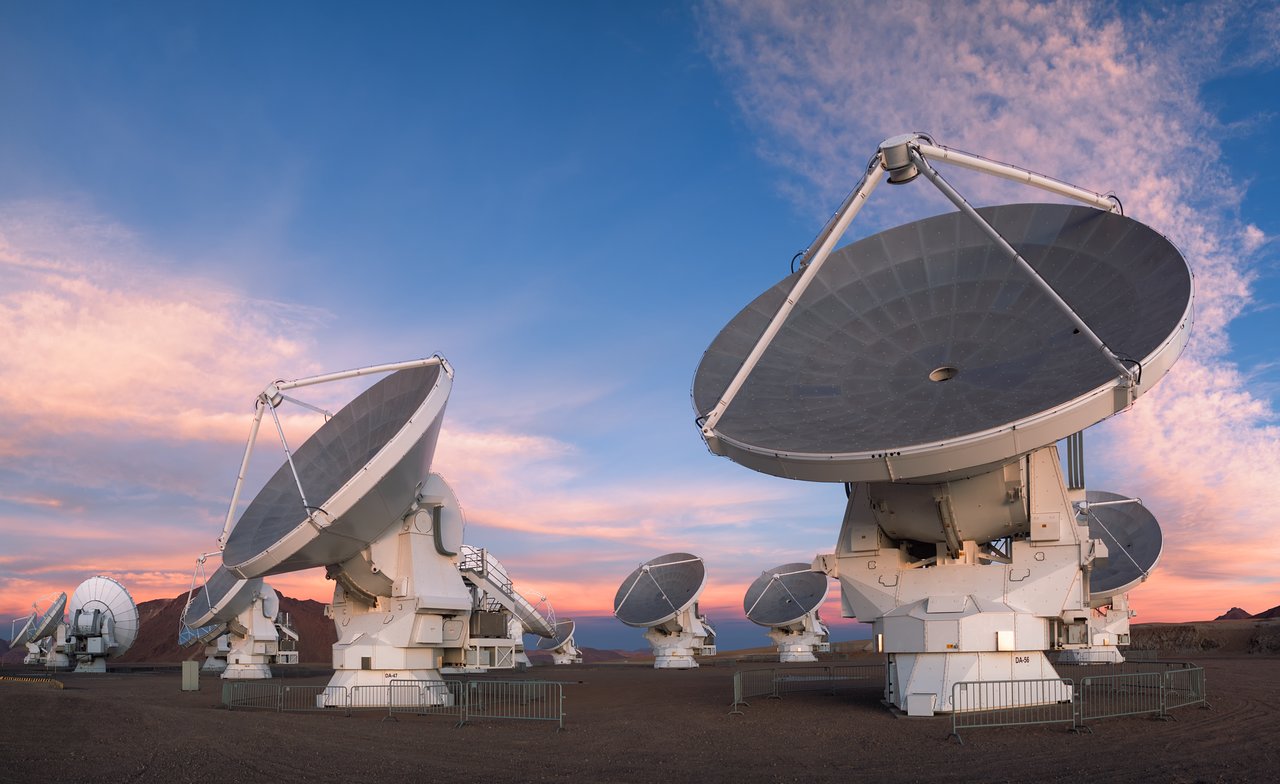}
\end{figure}

The Atacama Large Millimeter/submillimeter Array (ALMA) is an astronomical interferometer of 66 radio telescopes in the Atacama Desert of northern Chile, which observe electromagnetic radiation at millimeter and submillimeter wavelengths (Fig. \ref{alma}). The array has been constructed at the altitude of 5,000 m at the Chajnantor plateau -- near the Llano de Chajnantor Observatory and the Atacama Pathfinder Experiment. The site is perfect due to its high elevation and low humidity. The array has much higher sensitivity and higher resolution than earlier submillimeter telescopes such as the single-dish JCMT or existing interferometer networks such as the Submillimeter Array or the Institut de Radio Astronomie Millimétrique (IRAM) Plateau de Bure facility.

\begin{figure}[!t]
\caption{Spectrum of PH$_3$ $(1-0)$ with ALMA  \cite{greaves2020}}
\label{almas} 
\centering
\includegraphics[width=9cm, height=5cm]{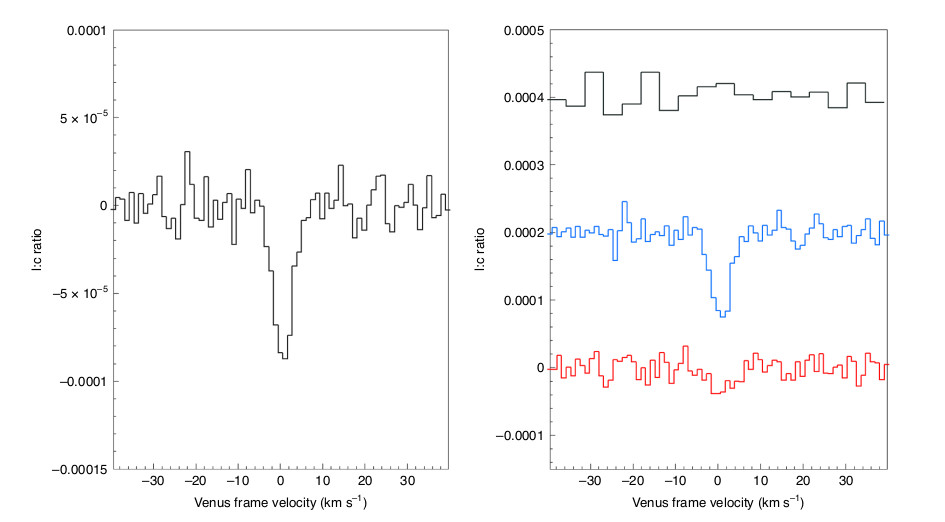}
\end{figure}

Figure~\ref{almas} shows the PH$_3$ ($1 - 0$) spectrum of the whole planet, with $1\sigma$ errors  of $0.11 \times 10^{-4}$ per 1.1 km/s spectral bin in the left panel. The right panel shows the spectra of the polar (histogram in black), mid-latitude (in blue) and equatorial (in red) zones on the planet.

A comparison of the production and destruction of phosphine and a better understanding of the temperature and pressure on Venus is an integral part of this study. 
The life time of phosphine in the layers at a height larger than 80 km is about $10^3$ s as it is destroyed by UV radiation.
 Near the base, due to collisions, the molecules are destroyed in $10^8$ s.
In mid-regions, there is no data about the lifetimes. It is expected to be about $10^3$ years. 
\cite{seager2020} proposed a hypothetical lifecycle in the Venusian atmosphere, the arial biosphere. Life could reside in the droplets, decicated spores that stay on higher levels of the atmosphere and when they reach the temperate zone, they get metabollically active, reproduce. They drift upwards and downwards as they evolve.

 Life as we know it, cannot survive in the presence of sulphric acid and the conditions in the Venusian atmosphere. However, possible life-forms can have an unknown photochemistry and hence the question is still open.

\section{The Controversy}

There are two big questions in this study. Firstly, does the signal detected correspond to phosphine and no other molecule?  Secondly, if it is phosphine, is it caused by life? 

To answer the first questions, the data has been multiply analysed by various research groups to verify the presence of phosphine. The next step would be to reobserve the signal repeatedly to see how it is distributed on Venus, does it change between day and night, or seasons, or different regions? {\em IF} it is truly related to life, then we would expect some kind of variation. And {\em IF} it truly is life, it won't produce only one such molecule, there  would be many more such molecules being produced by the complete ecosystem. 

The data obtained was very noisy and there were various algorithms were used to reduce the data and extract the signal without causing an artificial signal. Several observations were used, 18 months apart, to ensure that the signal was not spurious and applying different kinds of algorithms to identify the spectral line in multiple ways in repeated efforts by teams taking in inputs from experts and referees. It took almost 3 years from the first detection on JCMT to publication.

The authors of the discovery paper stated clearly that `Even if confirmed, we emphasize that the detection of PH$_3$ is not robust evidence for life, only for anomalous and unexplained chemistry.' \cite{greaves2020}.
       
Shortly after the announcement,  the organizing committee of the International Astronomical Union (IAU) Commission F3 on Astrobiology released a statement criticising Greaves' team for the  press coverage of their discovery. `It is an ethical duty for any scientist to communicate with the media and the public with great scientific rigour and to be careful not to overstate any interpretation which will be irretrievably picked up by the press,'  adding that the commission `would like to remind the relevant researchers that we need to understand how the press and the media behave before communicating with them'.

The IAU statement was condemned by many, including the commission’s own members, due to which the statement was retracted by the IAU executive. The IAU Executive stated  that the organizing committee of Commission F3 had `been contacted to retract their statement and to contact the scientific team with an apology'. 

Shortly after that, \cite{vill2020} contested the paper by \cite{greaves2020} and stated that the observed PH$_3$ feature with JCMT can be fully explained employing plausible mesospheric SO$_2$ abundances (~100 ppbv as per the SO$_2$ profile) and the identification of PH$_3$ in the ALMA data should be considered invalid due to severe baseline calibration issues. The team ended its abstract with the `suggestion' that Greaves' team retract its original paper – seen by some as unduly aggressive and hence an apology was made by Villanueva's team. `We agree that the sentence calling for retraction was inappropriate and we apologise for harm caused to the Greaves et al. team.' 

There were a few more criticisms of the paper by some groups like \cite{snellen2020,encre2020, thomp2020}. 
     
\cite{regreaves2020,souza2020} responded to the paper and reanalysed the data to recover PH$_3$ in Venus’ atmosphere with ALMA ($\approx 5\sigma$ confidence). They stated that the ALMA data are reconcilable with the JCMT  detection ($\approx$20 ppb) if there is order-of-magnitude temporal variation and more advanced processing  of the JCMT data is underway to check methods. They concluded that both ALMA and JCMT were  working at the limit of observatory capabilities and hence new spectra should be obtained as spectral ripples could potentially reducing the significance of real narrow spectral features. 

To add, a recent paper by \cite{mogul2020} re-examined  data obtained from the Pioneer-Venus Large Probe Neutral Mass Spectrometer (LNMS) from NASA’s Pioneer Venus spacecraft to search for evidence of phosphorus and confirmed the detection of phosphine on Venus.
 
 In the words of Lyman Beecher, `No great advance has ever been made in science, politics, or religion, without controversy'. We shall know.   

\section{The Future}
 
As of  now, the phosphine paper is the only set of results that has undergone peer review. Several papers have been submitted with a reanalysis of the data and if the findings are still under dispute, new data at different frequencies are required.

 Phosphine is difficult to detect from the ground but NASA’s airborne Stratospheric Observatory For Infrared Astronomy telescope, which flies at an altitude of over 13.7 km on board a modified Boeing 747SP, could confirm or deny the finding.   
 
Observations are also being planned in the Infrared using the NASA Infrared Telescope Facility (NASA IRTF) which is a 3-meter telescope in Hawaiii. Most telescopes are designed to look at faint sources. Venus is very bright and hence methods need to be planned to adapt to this bright source otherwise it saturates our detectors. JWST can look for such signals in far away planets. However, Venus is too bright for its detectors.

There will also be new missions for making in-situ measurements of the Venusian atmosphere to help build more realistic models.  The orbiter and atmospheric balloon mission Shukrayaan-1 by ISRO is under development and planned	under development and planned for 2023. The orbiter and lander  Venera-D by	Roscosmos is under development and planned for 2026.  NASA has proposed  a secondary payload VAMP for the  Venera-D lander.
 
 Small-scaled Venus mission could be launched in the near future to confirm the presence of phosphine and measure its  vertical distribution in the atmosphere. A whole series of missions could look for signs of life and even life itself. A golden era of Venus exploration lies ahead. 

The hypothesis of life in the clouds of Venus will gain scientific validity only after we have systematically ruled out all other chemical and geological processes that can explain the presence of phosphine on Venus and widened our explorations of this hostile, yet inviting twin.

\section{Conclusion}
The search of extraterrestial life has always been at the heart of human quest of Life, Universe and Everything. We have reached a stage where we are exploring planets of other Solar Systems and looking for habitability and signatures of life on them. Our Solar System is the immediate neighbourhood where we have started these explorations looking for indicators of water, habitability and life. The quest is not an easy one, however it is of great meaning and importance.
The exploration of Venus is one such attempt. Many more such attempts will be planned in the future and this can be a small step in the right direction in us, the universe, trying to find answers about itself.
Afterall, as Carl Sagan once said, `extraordinary claims require extraordinary evidence' (ECREE). That's what we are looking for.


\end{document}